\documentclass{cargese}\usepackage{graphicx}

\let\footnote\savefootnote
\let\footnotetext\savefootnotetext 
\let\footnoterule\savefootnoterule 
\normallatexbib

\def\cqg#1#2#3{Class. Quant. Grav {\bf{#1}} (#2) #3}
\def\jhep#1#2#3{JHEP {\bf{#1}} (#2) #3}

\begin{document}

\articletitle[Holography and IR Conformality in 2d]
{Holography and Infrared \\
Conformality in Two Dimensions}


\author{Marcus Berg}



\affil{Universit\`a di Roma ``Tor Vergata'' \\
via della Ricerca Scientifica, 1 \\
00133 Rome, Italy}    

\email{berg@roma2.infn.it}


\begin{abstract}

This is a very brief
review of some results from Refs. \cite{BerSam01} and \cite{BerSam02}.
In holographic renormalization, we
studied the RG flow of a 
2d $N=(4,4)$ CFT perturbed by a relevant operator,
flowing to a conformal fixed point in the IR. 
Here, the supergravity dual is displayed, and
the computation of 
correlators  is discussed.
The sample stress-energy correlator given here provides an opportunity
 to explicitly compare Zamolodchikov's $C$-function
to the proposal for a ``holographic $C$-function''.

\end{abstract}


First, I will recall how to compute correlators
holographically,
even in the presence of domain walls (for a review, see \cite{Skenderis02}). 

As a simple analogy to keep in mind, consider
a  medieval castle where soldiers are practicing cannon-firing
from atop the castle walls
into the interior courtyard. 
Let us say that when the wall has height $h_0$, 
a horizontally fired cannon ball hits the ground precisely in
the middle of the courtyard.
A priori the height of the wall $h$ and the angle of firing $\theta$
would of course be independent boundary conditions, but by requiring
that  the cannon
ball always hit the center of the courtyard,
these two become related: $h=h_0/\cos^2 \theta - r \tan \theta$, 
for a circular wall of radius $r$, and neglecting air resistance.
For an uneven castle wall, 
where the height of the wall varies as $h(x)$ with the distance $x$ around the
circumference of the castle, the same  relation relates $\theta(x)$ to
$h(x)$.
Going back to AdS/CFT, the requirement that a bulk field $\phi$ vanishes
in the deep interior relates a radial derivative 
at the boundary $\phi_{(1)}(x)$ 
to the boundary value $\phi_{(0)}(x)$.
This relation encodes the desired boundary correlator
(up to contributions from local counterterms):
\begin{equation}
\langle {\cal O}(x) {\cal O}(y) \rangle \quad \sim \quad
{\delta \phi_{(1)}(x) \over \delta \phi_{(0)}(y)}
\label{corr}
\end{equation}
and in fact, 
in the approximation of linear fluctuations around the domain wall,
we will have $\phi_{(1)}(x) = f(x) \phi_{(0)}(x) + {\cal O}(\phi_{(0)}^2)$,
and the correlator (\ref{corr}) {\it is} simply the relation $f(x)$
between the radial derivative and the boundary field,
enforced by the requirement that the field
vanish at a point in the deep interior.

To finish up this introductory part, I should admit that
apart from the more obvious deficiencies of the above mechanical analogy,
it is important that in general (for non-marginal
operators) the problem we are actually solving is
a {\it scaled} Dirichlet problem, such that one has to
factor out some radial dependence to obtain a finite boundary value.

To get to the main result presented in this contribution ---
a stress-energy tensor correlator ---
I need to briefly review the analytic domain wall solution
of~\cite{BerSam01} which interpolates between two AdS vacua. It was 
constructed as a solution of the three-dimensional 
Nicolai-Samtleben
gauged supergravity, in the case of 
16 supercharges and local $SO(4)\times SO(4)$
symmetry. This describes the $AdS_3\times S^3
\times S^3 \times S^1$ near-horizon geometry of the double D1-D5
system (references given in \cite{BerSam01,BerSam02}). 
The matter sector of this theory
consists of $n$ multiplets each containing $8$ scalars and $8$
fermions, and the nonpropagating fields are 
graviton, gravitini, and 
here also all $12$ vector fields. The $8n$ scalars parametrize the
coset manifold $SO(8,n)/(SO(8)\times SO(n))$.

The supergravity Lagrangian is given by
\begin{eqnarray*}
{\cal L}&=& {\textstyle \frac{1}{4}} \,\sqrt{G}R
+ {\cal L}_{\rm CS} + 
{\cal L}_{\rm kin}+
 \sqrt{G}\,V + {\cal L}_{\rm F}\;,
\label{L}
\end{eqnarray*}
where ${\cal L}_{\rm CS}$ is the Chern-Simons term for the vector fields,
${\cal L}_{\rm kin}$ is the kinetic term for the scalars, 
$V$ denotes the scalar potential and finally
${\cal L}_F$ contains the fermionic terms.
In addition to the local $SO(4)\times SO(4)$ gauge symmetry, the theory is
invariant under the rigid action of $SO(n)$, rotating
the matter multiplets. Specializing to $n = 4$ matter
multiplets, we 
consider a global invariance group 
$SO(4)_{\rm
inv}$ of the potential, embedded as the diagonal of the three
$SO(4)$ factors.

\begin{figure}
\begin{center}
\includegraphics[scale=.5]{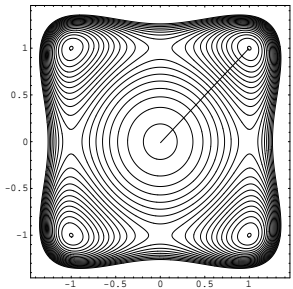}
\label{Vpic}
  \caption{\small 
Contour plot of the scalar potential $V(Z_1,Z_2)$, for $\alpha=1$ 
\cite{BerSam01}.
The supersymmetric flow from the central maximum to a local minimum is displayed. }
\end{center}
\end{figure}

Evaluation of the scalar potential $V$ on the two-dimensional space of
singlets under $SO(4)_{\rm inv}$ leads to
a potential plotted in fig.\ \ref{Vpic}.
Denote by $\alpha$ the
ratio between the couplings of
the two factors in $SO(4) \times SO(4)$.
The  potential depicted in fig.~\ref{Vpic}
is for $\alpha=1$. The figure
exhibits two inequivalent extremal points apart from the local maximum
at the origin. The saddle point at $(Z_1,Z_2)=(1,0)$ corresponds to a
nonsupersymmetric but stable vacuum, as was verified by direct
computation of the scalar fluctuations around this
point~\cite{BerSam01}. Here I will concentrate on the extremum located at
$(Z_1,Z_2)=(1,1)$, which preserves $N=(1,1)$ supersymmetry.
The flow to this point may be parametrized by a single scalar
$Z_1 = Z_2 =:~ (1/\sqrt{2})\sinh
(Q/\sqrt{2})$.

The ratio of the central charges of the dual conformal field theory at
this extremum and that of the CFT at the origin is given through the
Brown-Henneaux/Henningson-Skenderis relation
between central charge and AdS length, and becomes
\begin{equation}
{c_{\rm IR}}/{c_{\rm UV}} 
~=~ \sqrt{{V_{\rm UV}}/{V_{\rm IR}} } ~=~ 1/2
\; ,
\label{ccVV}
\end{equation}
supporting the conjecture that this point corresponds to a mass
deformation of the UV conformal field theory,
where half the degrees of freedom
are integrated out at long distances. The supergravity
spectrum around this point can be organized in $N=(1,1)$
supermultiplets, as expected.

With the standard domain wall ansatz
$ds^2~=~e^{2A(r)}\,\eta_{ij}\,dx^i dx^j + dr^2$,
and changing variables
for the domain-wall scalar to the convenient $y=\cosh \sqrt{2} Q$,
the first-order equations
admit the solution
\begin{equation}
\frac{(5-y)(y+1)^2}{16\,(y-1)^3} ~=~  e^{24gr} \; ,
\end{equation}
\begin{equation}
e^{6A(r)} ~=~ \frac{(5-y)^{4}}{128\,(y+1)(y-1)^6} \; .
\quad
\label{kink}
\end{equation}
This solution interpolates between the origin and the supersymmetric
extremum, preserving $N=(1,1)$ supersymmetry throughout the
flow. By studying linear fluctuations
around this analytic solution, one extracts holographic
correlation functions along the flow. 

Our fluctuations are not hypergeometric,
but can all be transformed 
to an equation known as the {\it biconfluent Heun equation}.
The two-point function of the boundary stress-energy tensor
in complex coordinates $z=x^1+i x^2$ is schematically
$\langle T_{z\bar{z}}T_{z\bar{z}}\rangle=
\delta g_{(2)} / \delta g_{(0)} + $ 
counterterms, and the result is
\begin{eqnarray}
\langle T_{z\bar{z}}(-p)T_{z\bar{z}}(p)\rangle &=& 
- {k|p|^4 \over 8}\left(\frac1{8|p|^2} +\frac{1}{2+8|p|^2 - 
\Psi_{-2}(p)}\right)  ,
\label{Tcorr}
\end{eqnarray}
where $\Psi_{-2}(p)$ is 
the Heun-function analogue of $\psi(p)$ for hypergeometric functions.
Unlike $\psi$, however, the coefficient $\Psi$ has no simple closed form,
but with some effort 
 the relevant asymptotics can be determined analytically.
Taking the infrared $p \rightarrow 0$ limit, the expression
(\ref{Tcorr}) has the appropriate CFT power-law behavior
$p^{2\Delta_{\rm IR}-2}$.
The correlator is the main result reported in this contribution.

A word about the factor $k$ in the correlator.
In the $N=(4,4)$ boundary conformal field theory,
$k$ denotes the equal levels of the two $SU(2)^2$
current algebras. In terms of the ten-dimensional 
double D1-D5
supergravity
solution with brane charge $N$, the appropriate 
relation is $k \sim N^2$.

Finally, the small distance
behavior of the Zamolodchikov $C$ function 
computed using this and similar correlators is
\begin{equation}
C_{\rm Zam} = 3k  \left( 1 - \frac1{4\sqrt{2}} \,|z| + {\cal O}(|z|^3)
\right) \;.
\end{equation}
This holographically computed
Zamolodchikov $C$-function can then be compared to the ``holographic $C$ function'',
which is simply given in terms of the superpotential $W$ as
$C_{\rm hol} = -3k/W(Q)$
where $Q$ is the scalar supporting the domain wall, as above.
The details of the comparison are left to the references.


\begin{acknowledgments}
It is my pleasure to thank H. Samtleben for collaboration. 
I would also like to thank the organizers of the Carg\`ese 2002 ASI.
This work was supported  by a Marie Curie Fellowship,
contract number HPMF-CT-2001-01311, and in part by
INFN, by the EC contract HPRN-CT-2000-00122, by the EC contract
HPRN-CT-2000-00148, by the INTAS contract 99-0-590 and by the MURST-COFIN
contract 2001-025492.

\end{acknowledgments}



\begin{chapthebibliography}{99}



\bibitem{Skenderis02}
K.~Skenderis, {Lecture Notes on Holographic Renormalization},
\cqg{19}{2002}{5849-5876}; hep-th/0209067.

\bibitem{BerSam01}
M.~Berg and H.~Samtleben, {An Exact Holographic RG Flow Between 2d
Conformal Fixed Points},  \jhep{05} {2002}{006}; hep-th/0112154.

\bibitem{BerSam02}
M.~Berg and H.~Samtleben, {Holographic Correlators
in a Flow to a Fixed Point}, hep-th/0209191.



\end{chapthebibliography}
\end{document}